\documentstyle[prl,aps,epsf]{revtex}         

\begin{document}
%
\twocolumn[\hsize\textwidth\columnwidth\hsize\csname @twocolumnfalse\endcsname 

\title{Reversible Formation of a Bose-Einstein Condensate}
\author{D.M.~Stamper-Kurn, H.-J.~Miesner, A.P.~Chikkatur, S. Inouye,
J. Stenger, and W.~Ketterle }
\address{Department of Physics and Research Laboratory of
Electronics, \\
Massachusetts Institute of Technology, Cambridge, MA 02139}
\date{ submitted} \maketitle

\begin{abstract}
We present a method of adiabatically changing the local phase-space density
of an ultracold gas, using a combination of magnetic and optical forces. 
Applying this method, we observe phase-space density increases in a
gas of sodium atoms by as much as 50-fold.
The transition to Bose-Einstein condensation was crossed reversibly,
attaining condensate fractions of up to 30\%.
Measurements of the condensate fraction reveal its reduction due to
interactions.
\end{abstract}
\vskip-1pc
]

 The physical properties of atomic gases change dramatically when quantum
 degeneracy is reached, i.e. when the ground state population approaches
 unity~\cite{grif:95}.
 Recent successes in reaching quantum degeneracy with Bose 
 gases~\cite{ande:95,davi:95,brad:97}
 have relied on non-adiabatic, irreversible methods such as laser and 
 evaporative cooling.
 The possibility of changing the ground state population by an adiabatic 
 change in the trapping potential had been overlooked for quite some 
 time~\cite{kett:92}.
 Indeed, in the case of an ideal gas, adiabatic changes 
 in the {\it strength} of the trapping  potential do not change the 
 ground state population~\cite{reif:65,footnote:1}.
 However, Pinkse and collaborators \cite{pink:97} recently
showed, both theoretically and experimentally,
that by changing the {\it form} of the trapping 
 potential,
the population in the ground state can be changed adiabatically.
For a non-degenerate gas the ground state population is identical to 
 the phase-space density $\Gamma = n \lambda_T^3$,
 where $n$ is the density of the gas and $\lambda_{T}$ is the thermal de
 Broglie wavelength.
 Within the type of trap deformations  considered in Ref.~\cite{pink:97}
 the maximum increase of phase-space density is limited to a factor of 20.
 
 In this Letter, we show that a more general deformation of the trapping
 potential can increase the phase-space density by an arbitrary factor, and we
 describe an implementation of  this scheme
 using a combination of magnetic and optical forces.  
 Furthermore, we demonstrate the ability to cross the Bose-Einstein 
 condensation (BEC) phase 
 transition and to change the condensate fraction of a Bose gas in a 
 reversible way.

{\bf Adiabatic increase in phase-space density}.
The type of trap deformations which we study can 
be understood with the following ``two-box'' model.
Consider a classical gas of $N$ atoms 
confined in a box of volume $V_{0}= V_{1} + V_{2}$
with an initial phase-space density $\Gamma_0$.
Suppose that the potential within a sub-volume $V_{2}$ of the box
is lowered to a final well-depth $U$.
In this final potential, the gas equilibrates at a temperature
$T$, and the density in $V_2$ will be higher than that in
$V_1$ by the Boltzmann factor $e^{U/k_B T}$.
Using the condition of adiabaticity and constant particle number, one
obtains the relative increase of
phase-space density in $V_{2}$ compared to
that in $V_{0}$ before compression:
\begin{equation}
        \ln \left(\Gamma_2/\Gamma_0\right) = \frac{U/k_B T}{1 + 
		(V_2 / V_1) e^{U/k_B T}}~.
        \label{classical}
\end{equation}
For deep potential wells, where $U/k_B
T \gg \ln(V_1/V_2)$, there is no increase in phase-space density since
all of the gas becomes confined in $V_2$, and  the
adiabatic deformation corresponds simply to a uniform compression of
the gas.
For shallow potential wells ($U/k_B T \ll \ln(V_1/V_2)$), the
phase-space density in $V_2$ increases as
$e^{U/k_B T}$.
As $U$ is varied between these limits, the phase-space density 
increase reaches a
maximum which is greater than $( V_1/V_2 )^{1/2}$.
Thus, by choosing an extreme ratio of volumes $V_1/V_{2}$, an
arbitrarily large
increase in phase-space density $\Gamma_2/\Gamma_0 $ is possible.

To demonstrate this phase-space density increase in a gas of trapped 
atoms, 
a narrow potential well (analogous to $V_{2}$) 
was added to a broad harmonic potential (corresponding to $V_{1}$) by
focusing a single infrared laserbeam at the center of a magnetic
trap.
First a gas of atomic sodium was evaporatively cooled to a 
temperature higher than the 
BEC phase transition temperture in the cloverleaf magnetic trap~\cite{mewe:96}.
The number of atoms and their temperature were adjusted by varying the
final radio frequency (rf) used in the rf-evaporative cooling 
stage~\cite{kett:96}. Afterwards, the magnetically
trapped cloud was decompressed by slowly reducing the currents
in the magnetic trapping coils.
Time-of-flight absorption imaging was used to characterize the cloud.
The total number of atoms $N$ was determined by integrating the column density 
across the cloud, and $T$ was determined by one-dimensional
Gaussian fits to the wings of the density distribution.
From these, we determined the fugacity $z$ of the gas by the relation
$g_3(z) = N (\hbar \bar\omega / k_B T)^3$,
and then its phase-space density by $\Gamma_0 = g_{3/2}(z)$,
where $g_{n}(z) =\sum \limits_{i=1}^{\infty} z^{i}/i^{n}$
\cite{footnote:2}.
Here, $\bar\omega$ is the geometric-mean trapping frequency of the 
magnetic trap, as determined by {\it in situ} 
measurements~\cite{stam:98b}. 
These phase-space density measurements were calibrated with images 
from magnetically trapped clouds at the phase transition.

The optical setup was similar to that used in Ref.~\cite{stam:98a}.
The infrared laser power was gradually ramped-up from zero to a power $P_c$
at which the onset of BEC was seen in time-of-flight images of 
clouds released from the 
deformed trap; this implied $\Gamma_f = g_{3/2}(1)=2.612$ 
for the final phase-space density.
The depth of the optical potential well was given
by $U_c/k_{B} = 37 \mu\text{K}~ P_c / w_{0}^{2}
~(\mu\text{m}^{2}/\text{mW})$,
where $w_{0}$ is the $1/e^{2}$ beam-waist radius at the focus. 
The ramp-up time was made long enough to ensure that the 
trap deformation was adiabatic, but also short enough to minimize 
heating and trap loss.
Ramp-up times of up to 10 s were used.
\begin{figure}[htbf]
\epsfxsize=70mm
\centerline{\epsfbox[0 0 595 293]{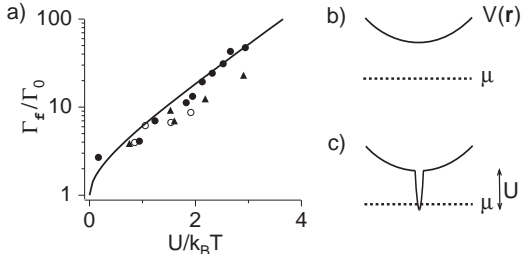}}
\caption{Phase-space density increase to reach BEC vs. normalized 
well-depth (a).
Various trap settings were used:
$\bar\omega = 2 \pi \times 100$ Hz, $w_0 = 9\, \mu\text{m}$ (triangles);
$\bar\omega = 2 \pi \times 100$ Hz, $w_0 = 18\, \mu\text{m}$ (open 
circles);
and $\bar\omega = 2 \pi \times 33$ Hz, $w_0 = 18\, \mu\text{m}$ 
(closed circles).
The solid line gives the prediction of Eq. (\ref{simple_theory_increase}).
(b) Prior to deformation, the harmonic trapping potential $V(r)$ 
holds a cloud above BEC the 
transition temperature ($\mu < 0$).
(c) When a potential well with depth
$U=-\mu$ is added, a small condensate forms. }
\label{fig1}
\end{figure}
Fig.\ \ref{fig1}a shows the increases in maximum phase-space density 
which were measured at three different settings of the trap parameters.
A maximum increase by a factor of 50 was obtained.
Condensates were observed in clouds with temperatures as high as 5 
$\mu$K.
Further increases were hindered by limitations in laser power
and by limits to the ramp-up time set by the various 
heating and loss processes in the deformed trap.
The scatter in the data is primarily due to statistical 
errors in our measurements of $U_c/k_B T$, at the level of 30\%, which 
arise from measurements of $P_c$ and $w_0$, and in
the determination of the transition point.

The well-depth $ U_{c} $ required to reach
BEC can be understood by a simple model depicted in  Fig.~\ref{fig1}.
We begin with a gas in a harmonic trap above the BEC
transition temperature, i.e. its chemical potential $\mu < 0$
(Fig.~\ref{fig1}b). 
Its initial phase-space density is given by $\Gamma_0 =
g_{3/2}(z) $,
where $z = e^{\mu/k_B T}$ is the fugacity.
In  Fig.~\ref{fig1}c,  a narrow potential with depth $U_{c}$ is added 
until $ U_c = - \mu$, at which point the gas begins to Bose condense.
The phase-space density thus reaches the critical value $g_{3/2}(1)$,
so that the increase in maximum phase-space density is given by
\begin{equation}
\frac{\Gamma_f}{\Gamma_0} = \frac{g_{3/2}(1)}{g_{3/2}(\exp({-U_c/k_B
T}))} \quad .
\label{simple_theory_increase}
\end{equation}
This prediction, shown in Fig.~\ref{fig1}a,
describes our data well, and accounts for the universal behaviour of
our measurements over a wide range of temperatures and 
well-depths.

Note that in this simple model we made the implicit assumption that 
the initial and final temperatures of the cloud were equal.
We can remove this assumption by considering instead the condition of 
constant entropy.
The entropy per non-condensed particle in a harmonically confined Bose gas
is determined uniquely by its fugacity $z$~\cite{pink:97}:
\begin{equation}
        {\frac{S}{N}}(z) = 4 \frac{g_4(z)}{g_3(z)} - \ln z \quad .
        \label{harmonic_spern}
\end{equation}
This equation describes the entropy of the 
gas before compression, with the fugacity 
given by $z_0$.
After compression, because of the small volume of the potential well, 
the entropy per particle  is approximately that of a harmonically 
trapped gas (Eq.\ (\ref{harmonic_spern})) with 
fugacity $z_f = e^{-U_c / k_B T}$.
Here $T$ is the {\it final} 
temperature of the gas, which is generally higher than the initial 
temperature~\cite{footnote:3}. Constant entropy then implies $z_0 = 
z_f$, and thus one obtains Eq.\ (\ref{simple_theory_increase}) where 
$T$ is now the final temperature, as we have plotted in Fig.~\ref{fig1}a.

One may also consider the process of adiabatically increasing the phase-space 
density as a change in the density of states $D(\epsilon)$ of the system.
By increasing the well-depth in a small region of the trap, we lowered only
the energy of the ground state and a few excited states. Thus $\Gamma$,
a local quantity,
is maximally increased as the ground state energy is brought ever closer to the
chemical potential, while the entropy, a global property of
the gas, is unchanged by the minimal modification of $D(\epsilon)$.

\begin{figure}[hhhh]
\epsfxsize=70mm
\centerline{\epsfbox[0 0 480 222]{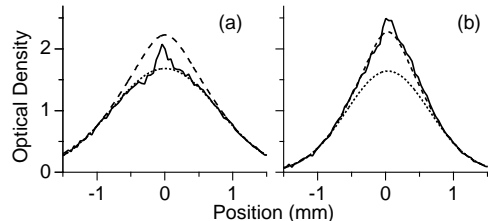}}
\caption{
Momentum distributions of the thermal cloud at the BEC phase 
transition show no Bose-enhancement for the deformed trap (a), but a 
clear Bose-enhancement for the purely magnetic trap (b).
Both distributions show a small condensate peak.
Lines show a Maxwell-Boltzmann distribution (dotted) and a
Bose-Einstein distribution (for $z =1$) (dashed) for clouds in a harmonic trap, 
which were fit to the thermal wings. The momentum distribution is 
shown as a profile across absorption images taken after 40 ms of 
ballistic expansion.
}
\label{fig2}
\end{figure}

The fact that global properties of the gas are not affected by the 
trap deformation can
also be seen in the momentum distributions probed by time-of-flight
imaging~\cite{mewe:96,ensh:96}.
The onset of BEC in the combined optical and magnetic trap is 
signaled only by the formation of a condensate peak.
The remaining thermal cloud is well fit by a Maxwell-Boltzmann 
distribution, which describes a magnetically trapped cloud far from 
condensation (Fig.~\ref{fig2}a).
In contrast, at the BEC transition in the harmonic magnetic trap, the 
momentum distribution of the thermal cloud is clearly Bose-enhanced at 
low momenta (Fig.~\ref{fig2}b).

{\bf Adiabatic condensation}. We now turn to the studies of adiabatic, 
i.e., reversible, condensate formation. 
A cloud of about $50 \times 10^6$ atoms was evaporatively cooled down 
to the the transition temperature in the magnetic trap,
at trap frequencies of  
 $\omega_r = 2 \pi \times 20\, \text{Hz}$ and $\omega_z = 2 \pi \times 13\,
\text{Hz}$ in the radial and axial direction, respectively.
The power of the infrared laser beam (of radius $w_0 = 20\, \mu\text{m}$),
was ramped up over 1~s and held at a constant power for
a dwell time of 1.5 s.
Condensate fractions as small as 1\% 
could be distinguished from the normal fraction by their anisotropic
expansion in time-of-flight images~\cite{ande:95,davi:95}. The condensate number $N_0$ was
determined by subtracting out the thermal cloud background using 
Gaussian fits to the thermal cloud in regions
where the condensate was clearly absent. 
\begin{figure}
\epsfxsize=70mm
\centerline{\epsfbox[ 0 0 417 234]{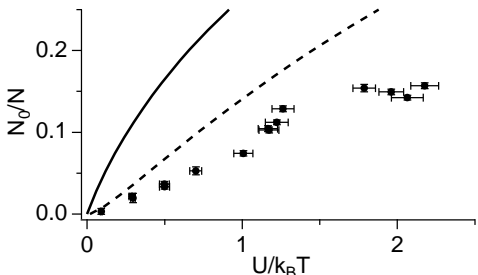}}
\caption{Condensate formation by adiabatic trap deformation. The condensate 
fraction is plotted against the (normalized) well-depth.
Lines show predictions for an ideal gas (solid) and for an interacting 
gas (dashed).}
\label{fig3}
\end{figure}
As shown in Fig.~3, the adiabatic trap deformation yielded
condensate fractions of up to 15\%;
we observed condensate fractions of 30\% with different trap settings.
By varying the dwell time, we confirmed that clouds for which
$U/ k_B T < 1.5$
suffered no significant losses of condensate number due to
three-body decay or heating, whereas those points with higher values of
$U / k_B T$ were affected by such losses.

For an ideal Bose gas, the result of this adiabatic change
can be understood as follows.
Before compression, the cloud of $N$ particles at the BEC transition
has an entropy $S_i$ given by
Eq.\ (\ref{harmonic_spern}) as $S_i = N \times 4 g_4(1) / g_3(1)$.
After compression, the situation is similar to that indicated in 
Fig.~1c, 
i.e. because of the small volume of the attractive well, the cloud
is well-described as a harmonically trapped
gas with $\mu = -U$.
Thus, accounting for the fact that condensate particles carry no 
entropy and again using Eq.\ (\ref{harmonic_spern}), which gives the 
entropy per {\it non-condensed} particle, we equate the 
entropy before and after compression and obtain
\begin{equation}
        \frac{N_0}{N} = 1 - \frac{4 g_4(1) / g_3(1)}{4 g_4(e^{-U/k_B T}) /
g_3(e^{-U/k_B T}) + U/k_B T}.
        \label{ideal_cf}
\end{equation}

However, this simple prediction does not
describe our findings well.
The theory described above has two  shortcomings.
First, the approximation  of 
using Eq.\ (\ref{harmonic_spern}) for the deformed trap is not strictly valid. 
However, calculations which accounted for the true shape of the deformed 
potential changed the prediction of
Eq.\ (\ref{ideal_cf}) only for $U/k_B T > 1$, and only slightly
improved the fit to our data.

A second shortcoming is the neglect of interactions.
It has been shown that in harmonic potentials, the condensate fraction
in a Bose gas with repulsive interactions
is reduced in comparison to that of an ideal gas~\cite{finite_t,nara:98}.
To estimate this effect in our system, we use 
the ``semi-ideal'' model of Ref.~\cite{nara:98}.
The thermal cloud is described as an ideal gas for which the chemical potential is 
raised by $ g n_{0} = 4\pi \hbar^{2} a n_{0} /m $, where $n_{0}$ is the 
maximum condensate density, $m$ the mass of sodium,
and $a = 2.75$~nm its scattering length 
\cite{ties:96}. This simply corresponds to using Eq.~(\ref{ideal_cf}) 
with the substitution $U \rightarrow U - g n_0$.
We determined $n_{0}$ using the Gross-Pitaevskii equation in the 
Thomas-Fermi limit \cite{baym:96}, and a harmonic approximation for the deformed 
trap potential at its center.

This approach predicts a significant reduction of the condensate 
fraction (Fig.~3, dashed line), and  the 
improved agreement with our data is strong evidence for this effect. 
In contrast to related studies  in  purely harmonic traps 
\cite{mewe:96,ensh:96}, which did not show evidence for interaction 
effects, 
this depletion is strongly enhanced by the  shape of the 
potential we are using.
The mean-field energy of the condensate $g n_0$ is large 
because the  condensate forms in the tight
optical potential, while the transition temperature $T_c$ is small since
it is determined by the weak magnetic potential.  

The reversibility of crossing the BEC phase transition was demonstrated
by preparing a magnetically trapped cloud just above $T_{c}$. We then 
sinusoidally modulated the power of the infrared light at 1~Hz, 
between zero and 7~mW. This modulation frequency was
significantly smaller than the magnetic trap frequencies
($\omega_{r} = 2 \pi\times 48~\text{Hz}$ and  $\omega_{z} = 2 \pi \times
16~\text{Hz}$). These low frequencies and a large optical focus
($w_{0}=18~\mu $m) were used to minimize trap loss due to inelastic 
collisions.

During the first seven condensation cycles the condensate
fraction oscillated between zero and 6\% (Fig.~4);
later probing
showed repeated condensation for at least 15 cycles.
The peak of these oscillations decreased slowly in time.
The temperature also oscillated, with an amplitude of about 100 nK,
while gradually rising by about 10 nK/s.
This heating and the decrease of the peak condensate fraction
are consistent with similar behaviour in clouds held at a constant
infrared power, which result from beam jitter,
spontaneous scattering, and three-body decay\cite{stam:98a}.
Thus, within the stability limitations of our optical setup, the
repeated crossing of the BEC phase transition appears fully adiabatic.
\begin{figure}[htbf]
\epsfxsize=60mm
\centerline{\epsfbox[0 0 375 373]{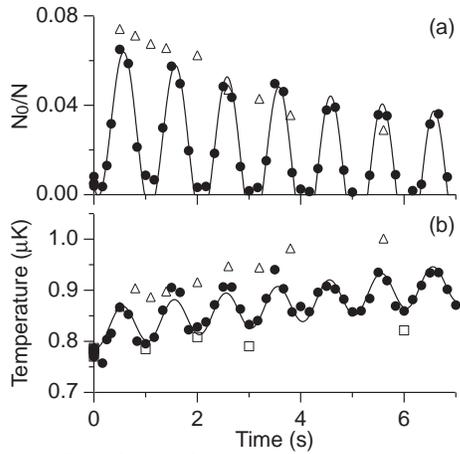}}
\caption{Adiabatic cycling through the phase transition. Shown is the
condensate fraction (a) and the temperature (b) vs time for the case
of a modulated infrared beam (closed circles), an infrared 
beam ramped up to a constant power
(open triangles), and no infrared light (open squares).
The solid lines are guides to the eye.}
\label{fig4}
\end{figure}

This method of creating condensates provides
insight into their formation, which was recently studied experimentally
\cite{mies:98} and theoretically 
 \cite{gard:97}.
For example, in the experiment described above, the condensate
fraction was found to lag about 70~ms behind the modulation of
the laser power, which is a measure for the formation time.

In other experiments, by switching on the infrared light instantly,
we  observed condensation on timescales
much faster than the oscillation periods in the magnetic trap and
along the weakly confining axis of the optical trap.
The resulting condensates showed striations in time-of-flight images,
indicating that the condensates formed into excited states of the
deformed potential.
Such studies of shock-condensation might give new insight into the
formation of quasi-condensates and condensation into excited 
states \cite{kaga:92,gard:98}.

In conclusion we have demonstrated the adiabatic Bose-Einstein
condensation of an ultracold gas of atomic sodium. 
Changes in the trapping potential resulted in 
large phase-space density 
increases and allowed for repeated crossings of the BEC phase transition. 
This method allows for detailed studies of condensate formation and the 
phase transition. 
The combined trapping potential widens the range of trap parameters 
over which BEC can be studied.
This was used to strikingly enhance the role of interactions, and 
led to higher transition temperatures (up to 5 $\mu$K) than achieved 
in purely magnetic traps.

We thank Michael R. Andrews for discussions.
This work was supported by the Office of Naval Research, NSF, Joint Services
Electronics Program (ARO), NASA, and the David and Lucile Packard Foundation. 
A.~P.~C.  acknowledges additional support from the NSF, D.~M.~S.~-K. from JSEP,
and J.~S. from the Alexander von Humboldt-Foundation.

\bibliographystyle{prsty}

\begin{thebibliography}{99}
\bibitem{footnote:3}
This approximation is valid for shallow potential wells
($U/k_B T \ll \ln(V_1/V_2)$), as discussed in the two-box model.
Numerical calculations confirmed its validity for all data in 
Fig.~\ref{fig1}a.
Note that the effective volume of 
the magnetic trap ($V_1$) increases with temperature.
\end{thebibliography}

\end{document}